# LONG-RANGE CORRELATIONS IN COVID-19 GROWTH

[1]Turaeva N., [2]Aripova N., [3]Oksengendler B.L.

[1]Webster University, Saint Louis, USA

[2]MICDS, Saint Louis, USA

[3]Arifov Institute of Ion-plasma and Laser Technologies, Tashkent, Uzbekistan

**Abstract**

The fractal statistics were applied to the daily new cases of COVID-19 in USA. The Hurst parameter, which indicates the long-range correlations in the growth, was calculated using a simple R/S method based on the fluctuations of the daily growth for several US states. The values of Hurst parameters for different states were analyzed using two controlling parameters, “stay-at-home” order and the population density.

## Introduction

The term fractal was first introduced by B. Mandelbrot [1] to describe geometrical objects with rough irregularities. In general, fractals are applied to objects in space and processes in time that can be characterized by properties of self-similarity, scaling, and fractal statistics [1-3]. Self-similarity means that smaller subunits of the fractals resemble the larger units, so fractals lack a single length (time) scale, and their structure (shape) is organized on multiple scales of length (time). The biological examples of spatial fractals are airways in the lung, vessels in the circulatory system, ducts in the liver, while temporal fractals are heartbeats, volumes of breaths, timing of the opening, and closing of ion channels.

The scaling property states that the values measured depend on the resolution used to make the measurement. Scaling relationships for correlations between measurements are often a power law that is a straight line on a logarithmic-logarithmic plot. For example, the quarter-power allometric scaling laws over different organisms are presented in nearly all biological rates and dimensions, including metabolic rate, lifespan, heart rate, DNA nucleotide substitution rate, length of aortas, cerebral gray matter [4]. Similarly, the statistics property states that the average or variance of a fractal object or process depends on the resolution used to make the measurement. As more data is included, the averages will continue to decrease or increase, so the real average does not exist. A number of different statistical properties have been measured to analyze fractal systems, which changes upon the resolution used to measure, including mean, variance, standard deviation, Fano factor, and rescaled range. Statistical properties of fractals are present in the timing of the action potentials in hearing nerve cells, in the spatial distribution of blood flow in the muscle of the heart, in the timing of the electrical activity of the heartbeat, or in the changing of electrical voltage across the cell membrane of white blood cells [2,3]. These studies show that fluctuations occur at all time scales, scale-invariant, and the time series exhibit long-range power-law correlations. This fractal notion was applied to interbeat interval fluctuations to distinguish certain disease states of the heart from the healthy states [5-7]. It is noteworthy that time series of the heart rate fluctuations for different heart conditions show nearly identical means and variances, suggesting no clinically relevant difference [5]. However, it was shown by using the method of detrended fluctuation analysis (DFA) that under normal conditions, beat-to-beat fluctuations in heart rate display long-range correlations typically exhibited by dynamical systems far from equilibrium, while heart rate time series from patients with severe congestive failure show a breakdown of such long-range correlation [7].

It is necessary to highlight the use of fractal concepts in the analysis of the electroencephalogram for the diagnosis of epilepsy [8]. The basic idea consists of the correspondence of epileptic seizures to the strict periodicity of the peak-wave signal, while the healthy state of the brain corresponds to a certain ratio of the order of chaos in the signal [9]. Moreover, the duration of the intermittency in the signals was studied by fractal analysis [10]. The fractal analysis method turned out to be so subtle that it differentiates the disease and health states in twin children [11].

Another major breakthrough achievement of the fractal analysis is in the diagnostics of diabetes. Three approaches have been proposed for glucose variability (GV) : one approach (A. Goldberger [12]) consists of the accordance of the fraction of the time, during which the system is out of the health corridor with the severity of the disease, i.e. when the blood sugar concentration was either higher (hyperglycemia) or lower (hypoglycemia) of the health corridor. The second diagnostic approach for GV belongs to B. Kovachev [13], who analyzed the statistics on up and down nonregular oscillations of glucose in glycemic curves. In the third approach for GV [14], the areas of glucose emissions are considered, which are defined as a product of the height of the peaks and their duration. It was shown that the distribution of glucose emissions is of a fractal character. In general, the last approach, similar to the approach developed by Hurst, effectively takes into account two other approaches developed by Goldberger [12] and Kovachev [13].

Thus, the exceptional adaptability of fractal analysis of both spatial objects and temporal processes in a number of nonlinear biomedical systems is completely established, and it is becoming increasingly clear that fractality is focused on fluctuations, not on average values [15-18]. Note here that it is historically well-known that the father of medicine Ibn Sina (known in the west as Avicenna) used the palpation of the pulse as a diagnostic tool to differentiate different diseases [19,20]. It is extremely interesting that he used deviations, but not average values for pulse fluctuations and had a phenomenal counter ability in it [19,20]. At the present, a deep understanding of how the study of the fractality in a system proceeds, has been achieved; at first, fractal parameters are calculated, and then the question of why the fractal dimension takes those values is answered. Such approaches are rigorously used by H.E. Stanley [16,21], and it is no doubt that it has huge methodological perspectives.

In this work, we will apply fractal statistics to spreading of COVID-19 by considering time series of fluctuations of new daily cases of the disease available through the international reference website, Worldometer. The novel coronavirus disease, labeled by the World Health Organization (WHO) as COVID-19, was first reported in Wuhan, China, on December 31, 2019. Compared to the previously identified coronaviruses such as the Severe Acute Respiratory Syndrome coronavirus (SARS-CoV) and the Middle East respiratory syndrome coronavirus (MERS-CoV), the fatality rate of COVID-19 caused by the third SARS-CoV-2 is substantially lower, but it is more transmissible, so far as it has spread to over 185 countries, and infected more than 2.79M people. (Worldometer, April 25).

We will use a R/S method to calculate the Hurst parameter, the value of which should indicate the presence of long-range correlations in the spreading of the virus among the population. We will study long-range order in the fluctuations of the new daily cases depending upon the two factors: “stay-at-home” order and density of population. At first, we will look at the origination of correlations in the system with large fluctuations, and then we will analyze the time series of new daily cases of COVID-19 in US for long-range correlations.

**Fluctuations in systems with correlations**

Let us consider the influence of fluctuations on macroscopic behavior of a system. There are two fundamental concepts in theory of probability and statistics, the law of large numbers and central limited theorem, which are valid for dynamic physical systems, the linear size of which is much larger than the average intermolecular size. According to the first law, the average of random variables ($X_1,X_2,X_3,...X_n$) converges to the expected value (m) as the number n tends to infinity, while the central limit theorem establishes that when independent random variables are added ($X=X_1+X_2+X_3+...X_n$), their normalized sum tends toward the normal distribution with variance equal to $\sigma^2/n$. These two results imply that the parameter of the variation defined by the standard deviation as $\delta = (\langle\delta X^2\rangle/\langle X\rangle^2)^{1/2}$ is proportional to $\sigma/\sqrt{n}$. This parameter tends to zero as n tends to infinity, and it is valid for different fundamental distribution functions, such as delta-function, Poisson, binomial or Gaussian, when the fluctuations of statistically independent random variables are smaller than the mean.

However, there are situations when the fluctuations are large, and the variance is larger than the squared mean $\delta > 1$, the macroscopic behavior of the system is impacted. It is important for such systems to take into account the fluctuations to explain the mechanism of emerging and sustaining space correlations. It was shown [22] that large deviations from the mean can be realized at simultaneous presence of both non-linear kinetics of chemical reactions and non-equilibrium conditions. This leads to the non-Poissonian distribution [22,23] of particles in volume and causes space correlations. In the systems where autocatalytic chemical reactions are conjugated with diffusion of the components providing the connection between the volume elements, the space correlations are defined by internal parameters, and basically are described by the correlation length according to the formula:

$$l_{corr} \approx \sqrt{\frac{D}{\gamma}} \tag{1}$$

Here D is the coefficient of diffusion of X component, $\gamma$ is the damping rate, which is related to the rate limiting step of the chemical reactions. Note that around the bifurcation point, the coefficient $\gamma$ tends to zero, and the correlation length increases infinitively. Thus, nonlinear chemical processes generate the fluctuations in the system, causing local correlations, and diffusion processes make these correlations extended in space, as a result long-range ordering takes place in space.

We suppose that the spreading of COVID-19 in different countries can be also analyzed based on the fluctuations of the growth of new daily coronavirus cases. We assume that autocatalytic reactions such as $A + X \rightarrow 2X$ can be applied to the process of infecting healthy people by the virus-carriers, and the transportation of people in the country is described by the diffusion coefficient. By considering the infecting process irreversible and taking into account the recovering process, we can expect correlations in the growth of new cases. We will study correlation effects in spreading of COVID-19 in USA by analyzing the time series of the daily growth of new cases before and after the “stay-at-home” order and in states with different population density.

**R/S analysis**

We considered the time series of the daily growth of COVID-19 new cases for several US states [24]. We constructed the graphs for the fluctuations of the daily growth with zero-mean and unit dispersion and the cumulative fluctuations (fig.1). Note, that the data is constantly changing, and we used the data taken as of the end of April.

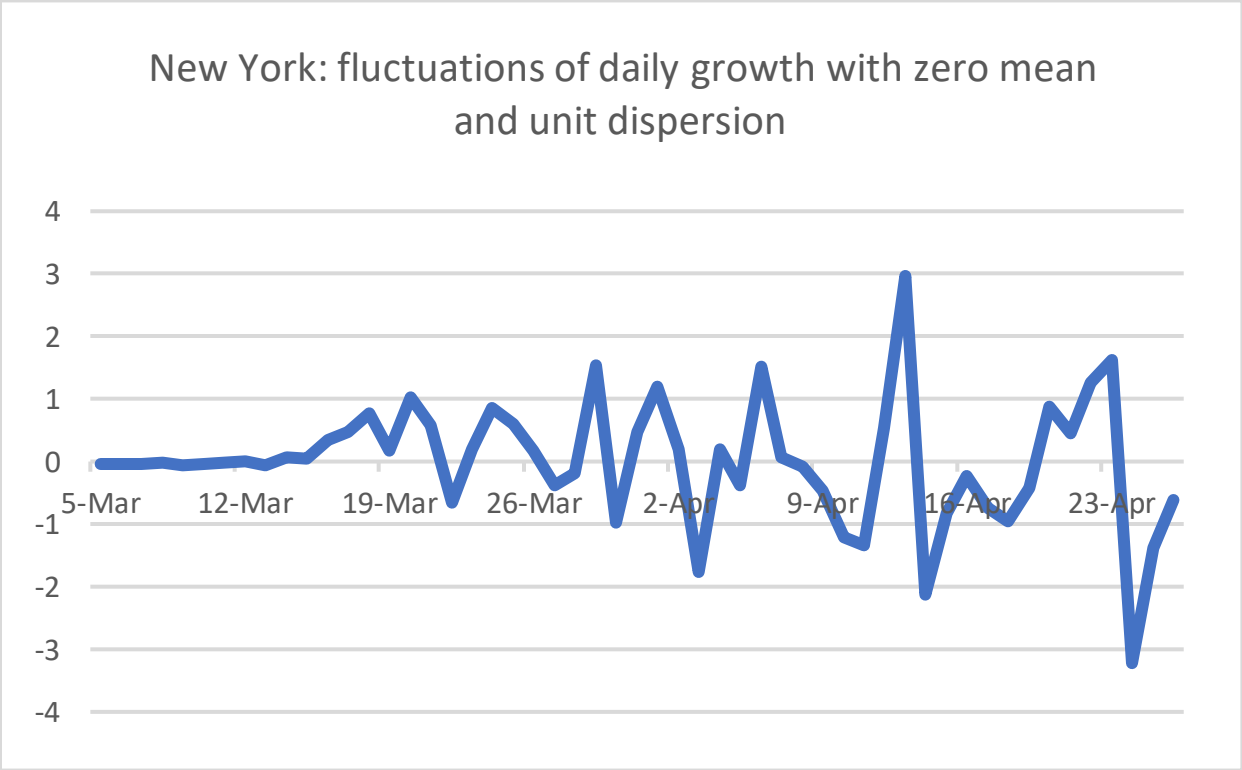

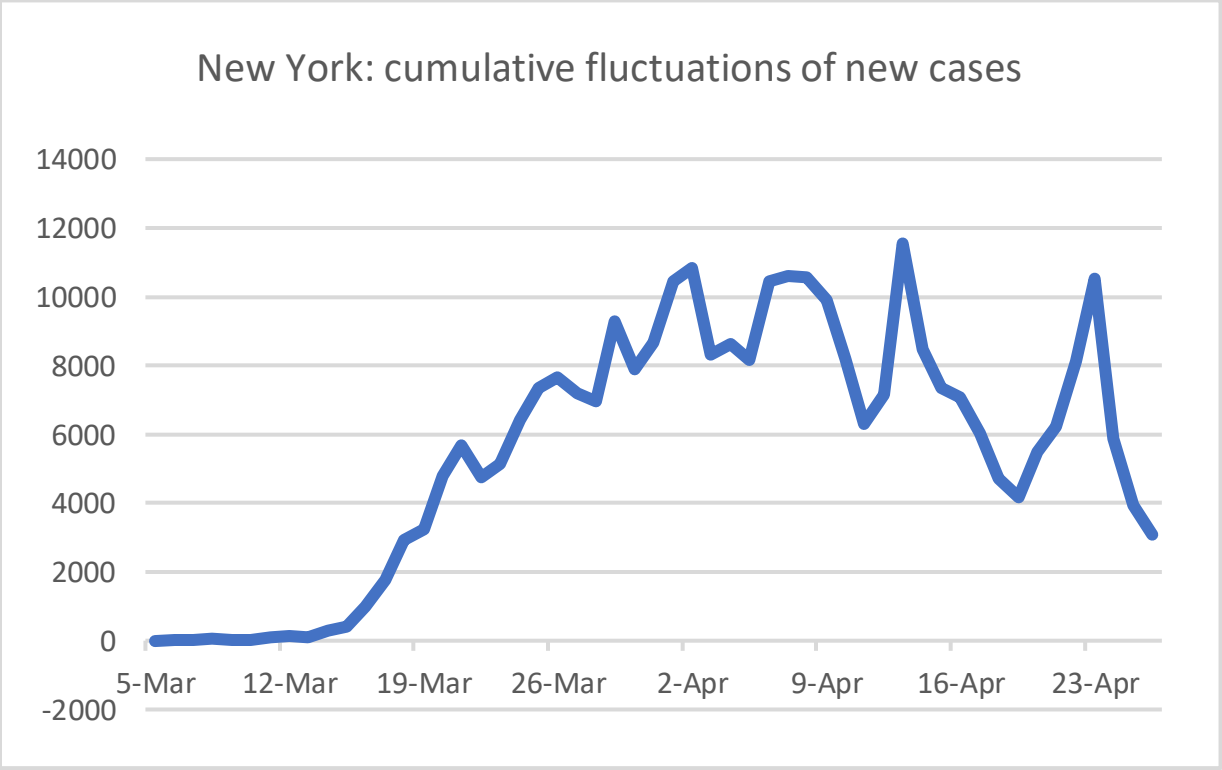

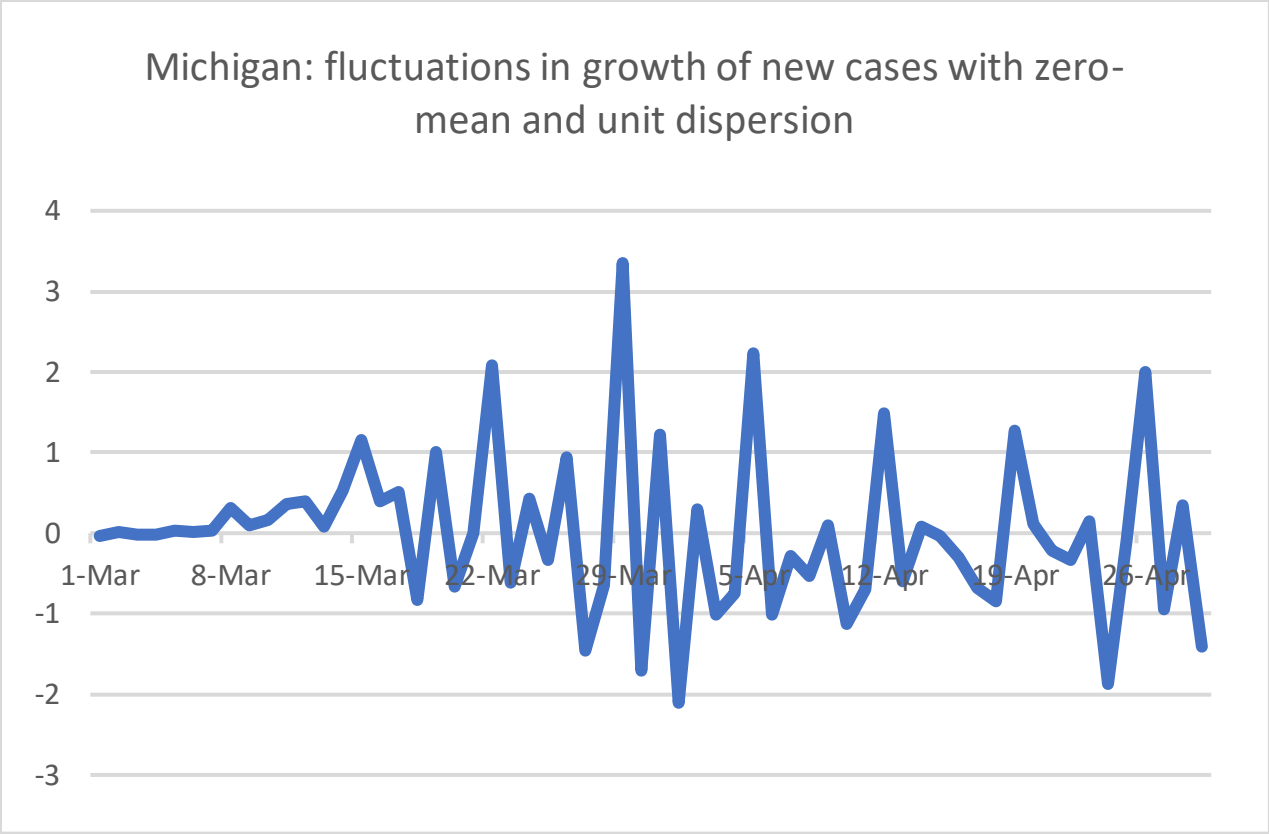
Michigan: fluctuations in growth of new cases with zero-mean and unit dispersion
4
3
2
1
0
-1
-2
-3
1-Mar
8-Mar
15-Mar
22-Mar
29-Mar
5-Apr
12-Apr
19-Apr
26-Apr

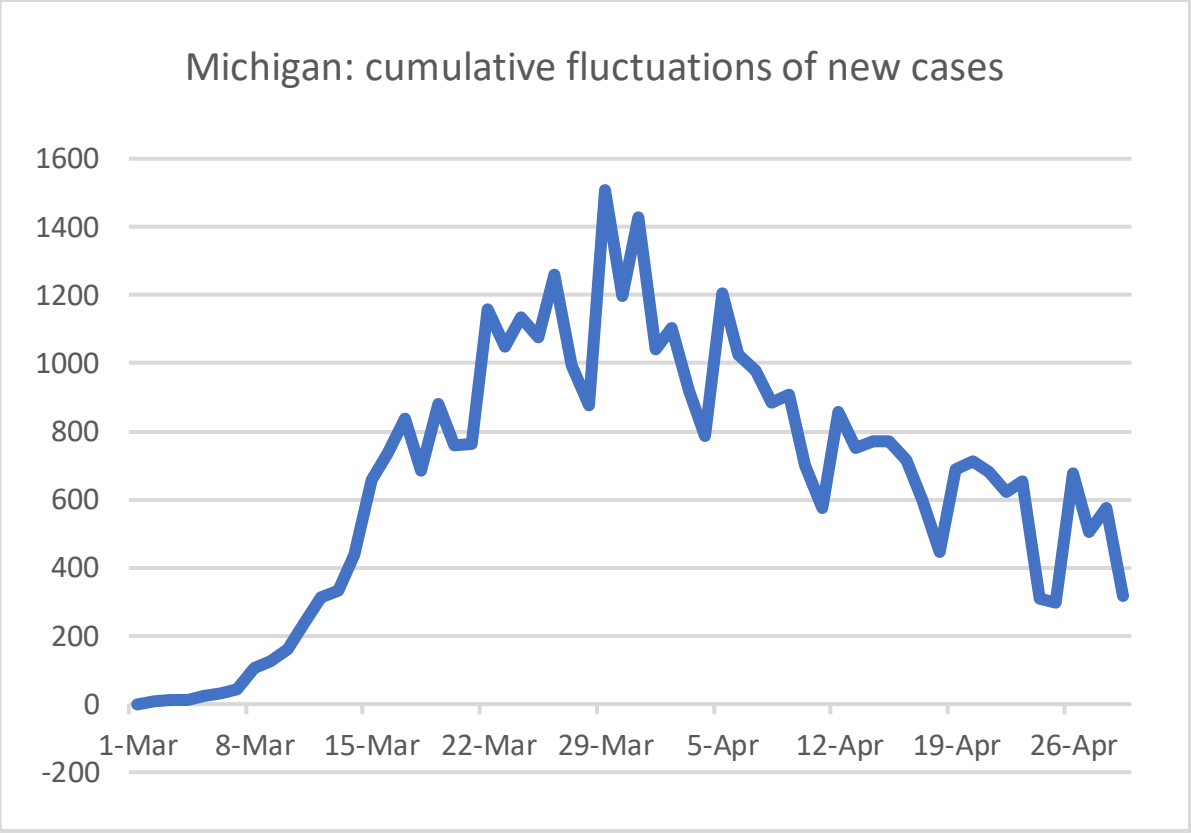

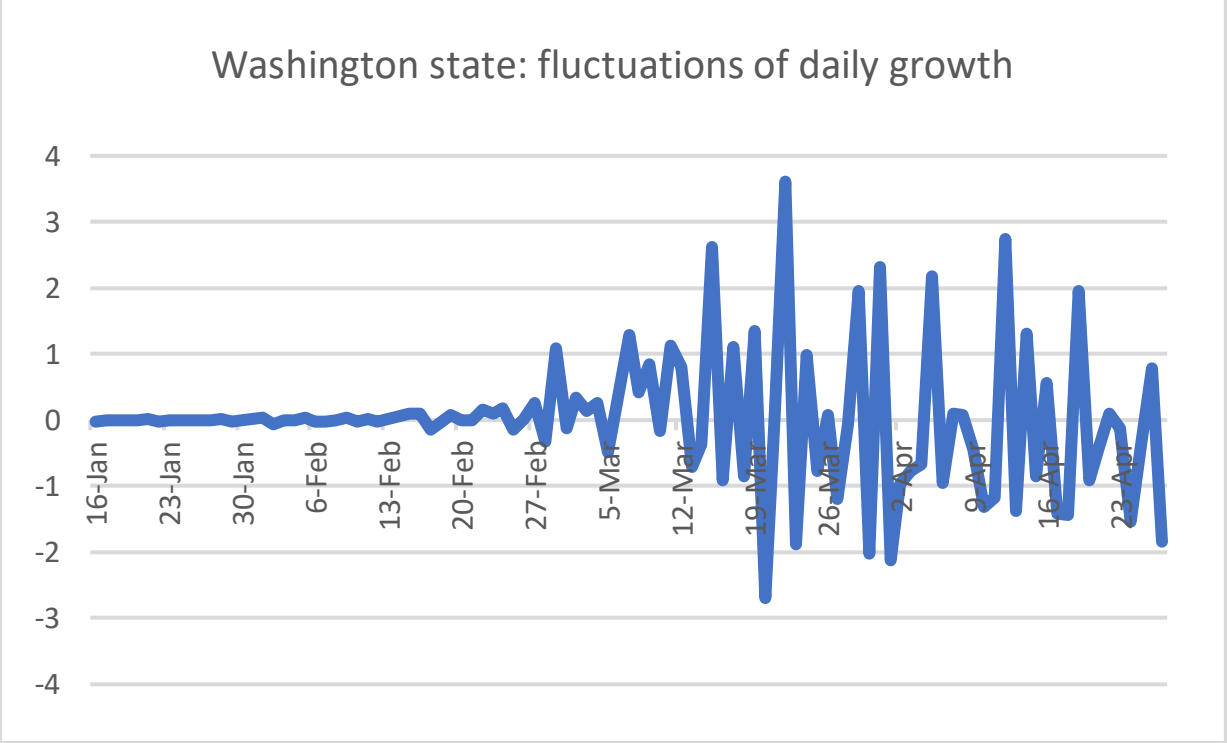

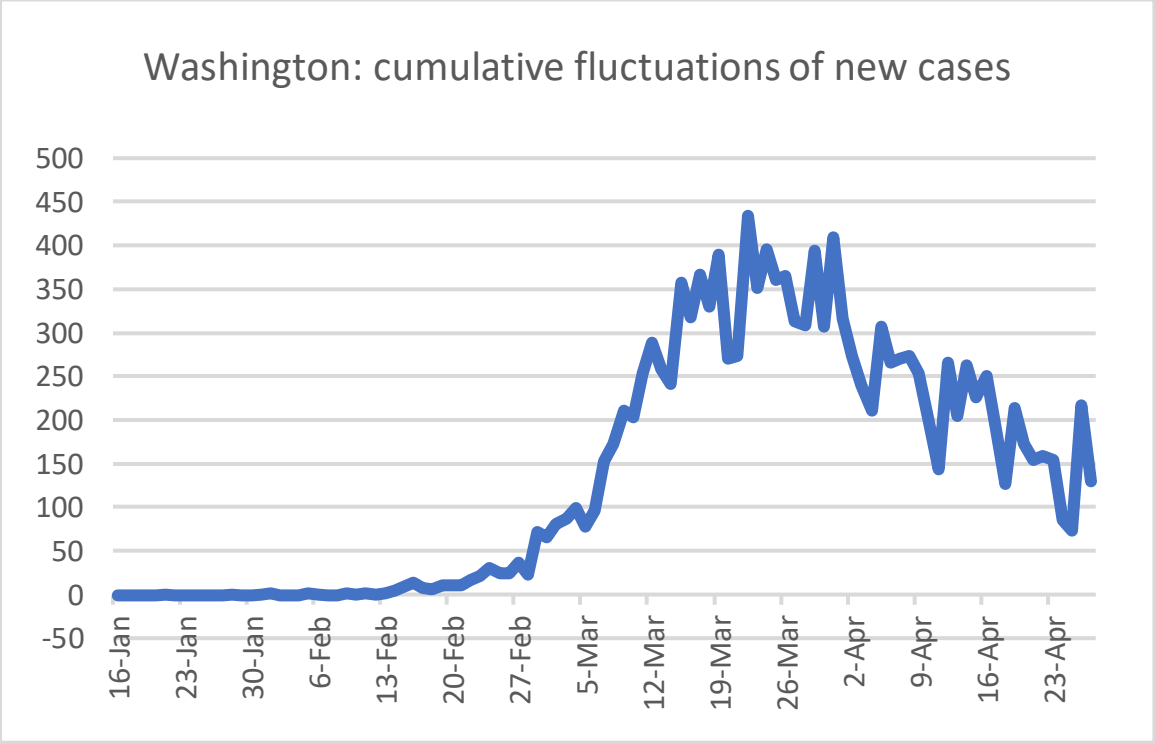

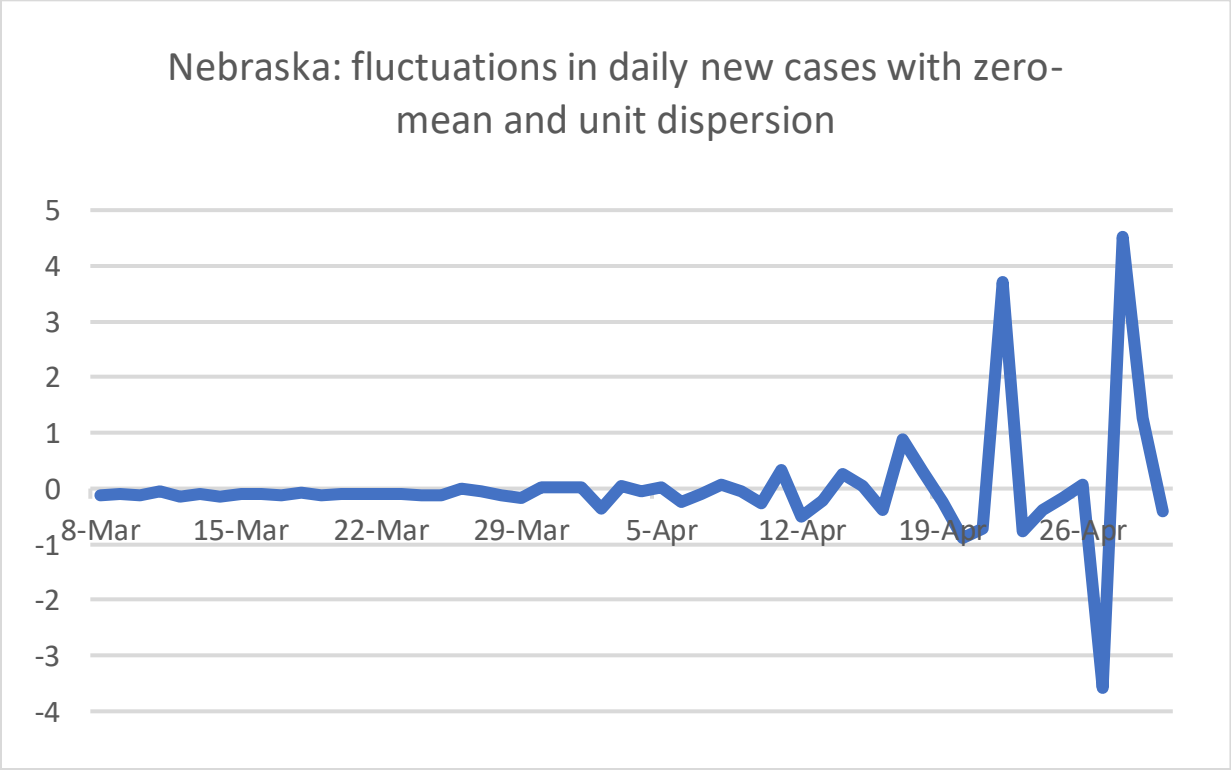
Nebraska: fluctuations in daily new cases with zero-mean and unit dispersion
5
4
3
2
1
0
-1
-2
-3
-4
8-Mar
15-Mar
22-Mar
29-Mar
5-Apr
12-Apr
19-Apr
26-Apr

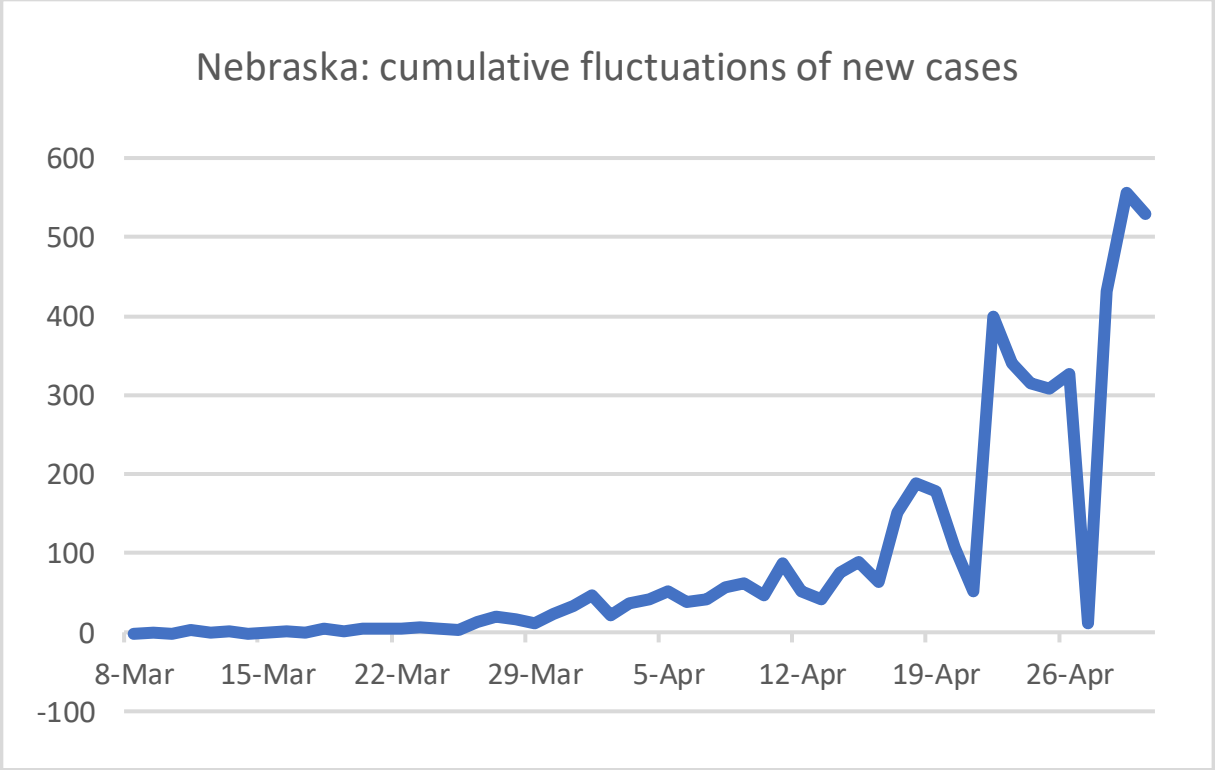

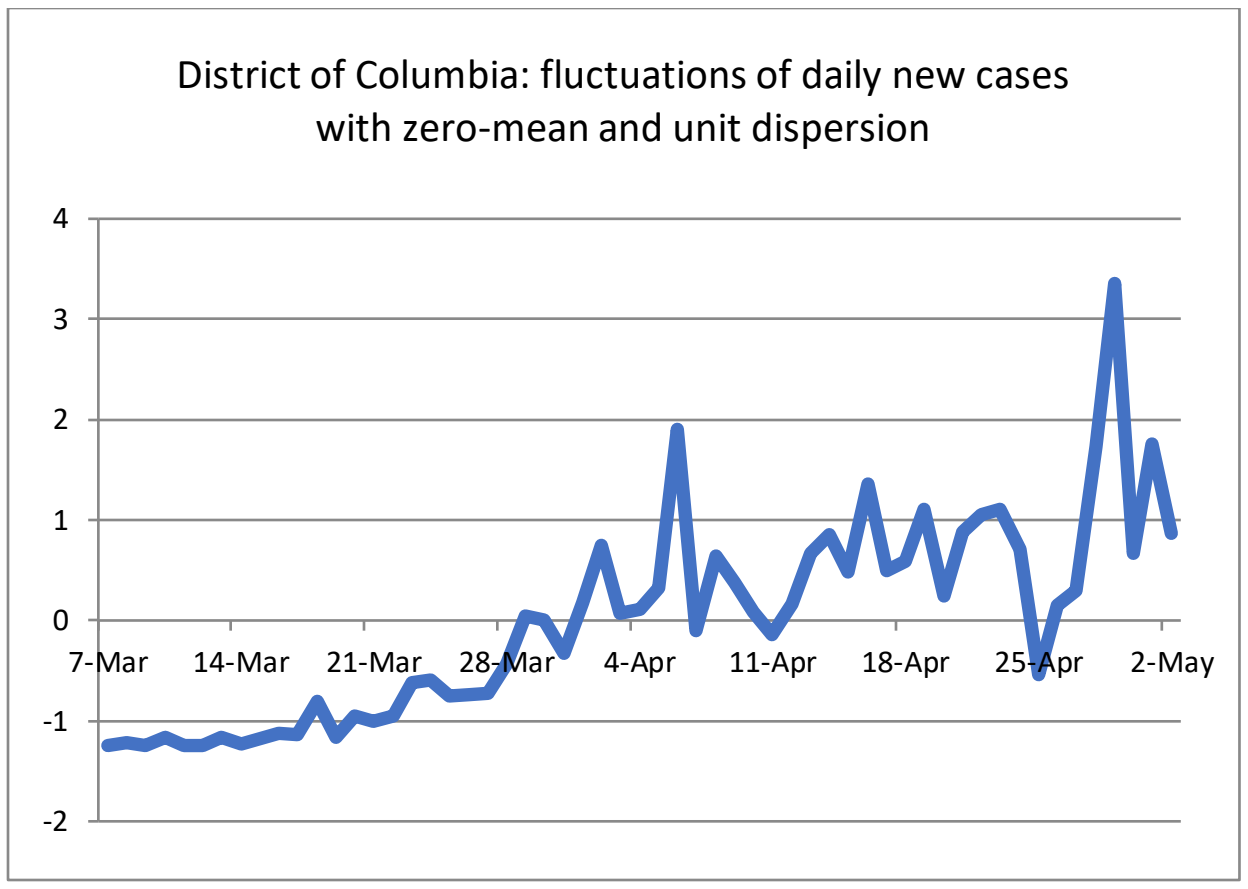

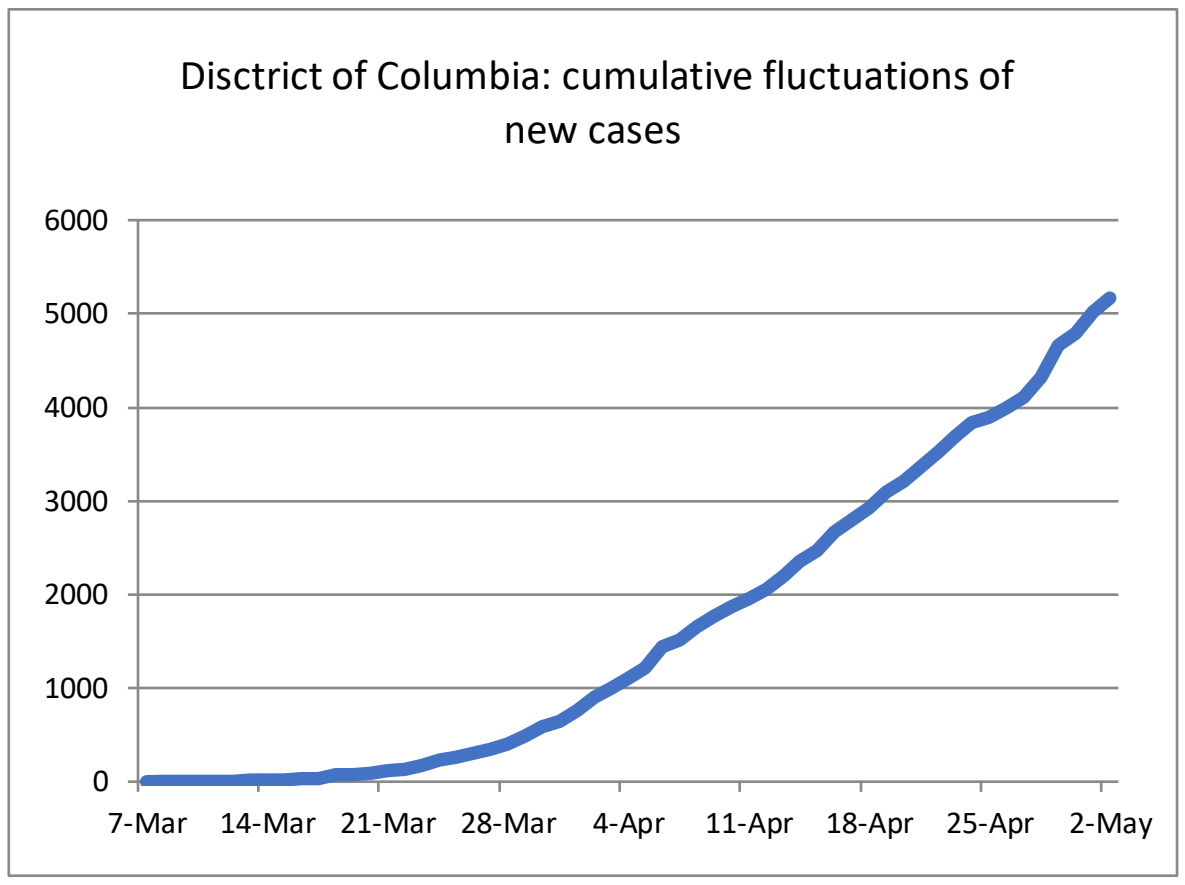

Fig 1. Fluctuations of daily new cases of Covid-19 with zero-mean and unit dispersion (left images) and the corresponding cumulative fluctuations for different US states (right images).

From Fig 1, it is seen that that cumulative fluctuations have different shapes for different states. The cumulative fluctuations have a tendency to increase sharply before the stay-at-home order in New York, Washington and Michigan states, which was announced around March 23, and are stabilized after the order, even decreasing in Washington and Michigan states. In Nebraska, the stabilization of the cumulative fluctuations in growth is not observed yet.

To quantify these considerations, we used a rescaled (R/S) method, allowing to calculate the Hurst parameter. The Hurst parameter is a useful statistical quantity for characterizing the long-term memory of time series. It quantifies the tendency of the time series to regress strongly to the mean or to cluster in a direction. Its value in the range 0.5 – 1 indicates a time series with long-term positive autocorrelation, meaning that a high value in the series will probably be followed by another high value, and that the values a long time into the future will also tend to be high. A value in the range 0-0.5 indicates a time series with long-term switching between high and low values in adjacent pairs, meaning the tendency to switch between high and low values will last a long time into the future. A value of 0.5 indicates a completely uncorrelated series.

We calculated the Hurst parameter for each state before and after the “stay-at-home” order. Due to the small number of available data (< 50), the values of the parameter are not very accurate. In this work, we show only a tendency of correlations in the growth of new cases for the change. The results are indicated in the table below. Note that there is no “stay-at-home” order in Nebraska although some restrictions have been put in place.

**Table 1. The Hurst parameter of the growth of Covid-19 daily cases before and after the "Stay-at-home" order**

| US states | before | after |
|---|---|---|
| New York | 1.17 | 0.45 |
| Washington | 0.62 | 0.26 |
| Michigan | 0.52 | 0.13 |
| Nebraska | 0.60 | |
| District of Columbia | 1.05 | 0.94 |

It is seen from Table 1 that the "stay-at-home" order is an important parameter which controls the persistency of the growth. Before the order in all states the persistency to the growth of new cases was observed, except the Michigan state which shows no correlation in the observations. However, after the order the Hurst parameter sharply decreases, except the District of Columbia. This change takes place mostly due to the restrictions of transportation of people, thus localizing them at home which corresponds to the small (almost zero) diffusion coefficient (see Eqn. 1). We can also compare the Hurst parameters of different states as a function of the population density. Due to the highest population density in District of Columbia and New York city, their Hurst parameters before the order are higher than that of other states showing high correlations in the observations.

As we noted above, the values of Hurst parameter are not very accurate because of the small number of data (<50) available so far. One more factor which could be responsible for inaccuracy of the results is the fact that it is not well known how many people in states have been infected to date since most people with Covid-19 experience mild illness or without any symptoms. These values can be improved in our further work as more data on new cases of Covid-19 become available.

## Conclusion

In this work, we have showed the presence of correlations in fluctuations of the growth of Covid-19 daily cases in several states of USA. As controlling factors, we considered two factors: the "stay-at-home" order and the population density of the states. It was shown that the Hurst parameter is a good indicator of decreasing the correlations in the spread of the disease due to the order. In all states with “stay-at-home” order, the parameter changes its value from persistency to antipersistency. The closer the value to zero, the stronger is the tendency for the fluctuations to revert to its long-term means value. The population density seems to be another crucial factor for spreading the disease due to the higher probability of the contingency of the people. In general, the dependence of the Hurst parameter on different controlling parameters such as “stay-at-home” order or the population density can be described in the framework of catastrophe theory [25].

Thus, the fractal methods of analysis are useful tools in the interdisciplinary biomedical sciences with a complex combination of biology, medicine, physics, technology, sociology. The fractal concepts can be also applied to modern ecology and economics, energy, and psychology. All of the examples mentioned above demonstrate a unique adequacy of fractal ideas. This pushed scientists to designate general laws and methods of the analysis, including fractal science, into a special global science of Complexity [26] which studies living matters, in which the complexity is particularly prominent. Epidemics can be also attributed to a series of complexity phenomena where biology, medicine, social sciences, ecology, and mathematics will be in one row. Therefore, more detailed information on the origination of collective processes can be derived from the fractal knowledge, where both local and nonlocal properties determine the correlation at different time and spatial scales.

**References**

1. B.B. Mandelbrot: The fractal geometry of Nature. New York, Freeman, 1982.
2. J. Feder: Fractals, New York, Plenum, 1988.
3. L. S. Liebovitch, Fractals and Chaos, Simplified for the Life Sciences, Oxford University Press, New York, Oxford, 1998.
4. G. West, and J.H. Brown, The origin of allometric scaling laws in biology from genomes to ecosystems: towards a quantitative unifying theory of biological structure and organization, The journal of Experimental Biology 208, 1575-1592 (2005).
5. A.L. Goldberger, Fractal Mechanisms in the electrophysiology of the heart, IEEE Engineering in medicine and biology, July, 47-52 (1992).

6. A.L. Goldenberger, D.R. Ridney, B.J. West, Chaos and fractals in human physiology. Sci Am., 262, 42-49 (1990).
7. C.K. Peng, S. Havlin, H.E. Stanley, and A.L. Goldberger, Quantification of scaling exponents and crossover phenomena in nonstationary heartbeat time series, Chaos 5, 82-87 (1995).
8. H. Haken: Principles of Brain Functioning.Springer-Verlag-Berlin-Heidelberg,1996.
9. H. Schuster: Deterministic Chaos. Physik-Verlag,1984.
10. P. Berge,Y. Pomeau, Ch.Vidal L'Ordre dans le chaos.Hermann,1988.
11. O. Sanafeeva, Differential diagnostic and prognostic values of EEG vegetological examination in febrile and epileptic paroxysms in children and adolescents. PhD Thesis, Tashkent, 24 (1995).
12. T.Danne, R.Nimri,T.Battelino, et.al. International consensus on use of continuous glucose monitoring. Diabetes Care. 2017;40:1631-1640.
13. B. Kovachev, C. Cobelli: Glucose variability: timing, risk analysis in Diabetes. Diabetes Care 39: 502-510 (2016).
14. B.L. Oksengendler, N.N. Turaeva, A.Kh. Ashirmetov, et.al. Nanofractals, Their Properties and Applications. In Horizons in World Physics Vol.292, P.1-35, 2019, New York:, Nova Science Publishers Inc.
15. G. Losa: The Fractal Geometry of Life. Revista di biologia Bioilogy Forum 102: 29-60 (2002).
16. J. Bassingthwaighte, L. Liebovitch, B. West Fractal Physiology Springer, NY,1994.
17. Z. Czemicki, W. Klonovski, L. Leibovich: Why Nonlinear Biomedical Physics? Nonlin. Biomed. Phys. July, 1-19 (2007).
18. K. Lambrini Vanabi, M. Atounty: Applications of fractals in medicine. Ann. of Univer. Craiowa, Math., Comp. Ser. 42:1,167-174 (2015).
19. Mones Abu-Asab, Hakima Amri, Marc S. Micozzi: Avicenna's Medicine: A New Translation of the 11th-Century Canon with Practical Applications for Integrative Health Care, Inner Traditions/Bear, 2013, 480 p.
20. Muhamed Azimov, The life and teachings of Ibn Sina, Indian Journal of History of Science, 21(3): 220-243 (1986).

21. S. Halvin, S. Buldyrev, A. Goldberger, H. Stanley et.al.: Fractals in Biology and Medicine. Chaos, Solitons, Fractals, 6:171-201,1995.
22. G. Nicolis, I. Progogine, Exploring Complexity: An introduction. New York, NY: W.H. Freeman, 1989.
23. H. Haken, Synergetics. An introduction: Nonequilibrium Phase Transitions and Self-organization in Physics, Chemistry, and Biology Springer, New York, 1983.
24. Coronavirus cases in the United States. (n.d.). Retrieved April 27, 2020, from https://www.worldometers.info/coronavirus/country/us/
25. R. Gilmore: Catastrophe theory for scientists and engineers. Dover Publications, 1993, 666 pp.
26. J. Nicolis: Dynamics of Hierarchical Systems Springer-Verlag-Berlin-Heidelberg,1986.